\documentclass[10pt]{article}

\usepackage{amsfonts}

\def\form#1{(\ref{#1})}

\def\s{\sigma}
\def\g{\gamma}
\def\al{\alpha}
\def\b{\beta}

 \def\B{{\cal B}} 
\def\H{{\mathcal H}}                   

\def\E{{\cal E}}

\def\ram{\mathop{\longrightarrow}\limits}
\def\um{\mathop{=}\limits}

\def\CADM{\mathop{\scriptscriptstyle \rm CADM}\nolimits}
\def\Kom{\mathop{\rm Kom}\nolimits}

\def\dim{\mathop{\rm dim}\nolimits}

\def\Re{I\kern-.36em R}                     

\newcommand{\be}{\begin{equation}}
\newcommand{\ee}{\end{equation}}
\newcommand{\ba}{\begin{eqnarray}}
\newcommand{\ea}{\end{eqnarray}}
\newcommand{\baa}{\begin{eqnarray*}}
\newcommand{\eaa}{\end{eqnarray*}}
\newcommand{\bat}{\be\left\{\begin{array}{l}}
\newcommand{\eat}{\end{array}\right.\ee}

\def\QDE{\rule{2.5mm}{2.5mm}}
\def\CVD{$\phantom{'}$\hfill\QDE}

\newtheorem{Theorem}{Theorem}[section]

\newtheorem{Remark}[Theorem]{Remark} %

\title{Boundary Conditions,  Energies and Gravitational Heat in General Relativity\\
\small 
 (a Classical Analysis)
}
\author{
M.\ Francaviglia\thanks{E-mail:
francaviglia@dm.unito.it}, M.\ Raiteri\thanks{E-mail:
raiteri@dm.unito.it}
\\
Dipartimento di Matematica, Universit\`a degli
Studi di Torino,\\
Via Carlo Alberto 10, 10123 Torino, Italy }

\date{}

\begin{document}

\maketitle


\begin{abstract}
The variation of the energy for a gravitational system is directly defined from the
Hamiltonian  field equations of  General Relativity. When the variation of the energy is
written in a covariant form it splits into two (covariant) contributions: one of them is the
Komar energy, while the other is the so--called covariant ADM correction term. When
specific boundary conditions are analyzed one sees that  the  Komar energy 
is related to  the gravitational heat while the ADM correction term plays the role of the
Helmholtz free energy. These properties allow to establish, inside a classical geometric
framework,    a formal  analogy between  gravitation and 
the laws governing the evolution of a thermodynamical system.
 The analogy  applies to stationary spacetimes admitting
multiple causal horizons as well as to AdS Taub--bolt solutions.\\


\end{abstract}
\section{Introduction}
One of the best--known classical methods to derive the gravitational
Hamiltonian  from the Hilbert Lagrangian is based on Noether theorem
\cite{Rosen,Trautman};  it relies on the  identification of the  Hamiltonian
density  with the Noether charge density associated with a timelike vector field $\xi$. The
Hamiltonian is then  obtained after integration of the Noether current on a
(portion) of a Cauchy hypersurface $\Sigma$ in spacetime. It is also
well--known \cite{libroFF,Lagrange,Robutti,KatzBicak,Trautman}  that the Hamiltonian 
obtained in this way splits into  a volume integral on
$\Sigma$ (which turns out to be  vanishing on shell being
proportional to the Hamiltonian constraints)  and a surface integral on 
$\partial\Sigma$ which is
nothing but the integral of the Noether superpotential, i.e. the
so--called  Komar potential \cite{Komar,Waldbook}. Therefore the gravitational energy,
defined as the  value on--shell of the Hamiltonian, just reduces to the integral of the
Komar form.  This simple definition of energy suffers however of at least one
remarkable drawback. Already when we deal   with simple solutions, e.g.\ 
asymptotically flat stationary spacetimes, it is well known that 
Noether techniques allow one to
obtain the expected value for the angular momentum but only one--half of the
ADM mass. Accordingly, the energy is affected by the so--called anomalous
factor
$1/2$ \cite{Katz}. A way to overcome the problem was suggested in
\cite{Lagrange,Katz,KatzBicak} (and references therein) and it basically consists in
correcting the Komar potential by  adding  to it a new (covariant) background--dependent
term  which allows to
\emph{cure} the anomalous factor. The final expression we end up  with is nothing but the
Noether superpotential associated to the so--called \emph{first order
covariant Lagrangian}
\cite{Nester,BYOur,libroFF,firstorder,Lagrange,Katz,KatzBicak},
obtained from the  Hilbert Lagrangian through the addition of a divergence term with an
arbitrary non--dynamical  background dependence.  

The technique of introducing a suitable background
dependence into the Lagrangian and the superpotential has been shown to work also for
spacetime solutions other than the  asymptotically flat ones and for generic spacetime
dimensions (see
\cite{MannLo,Remarks,BTZ,TaubBolt,libroFF,KatzBicak,KatzLerer,MannRa})  and this naturally
leads to the interpretation of the corresponding Noether energy as the   energy of each
solution
\emph{relative} to the chosen background.\footnote{
Of course, a number of physically  reasonable 
 geometric definitions  of gravitational energy other than the one based on Noether
theorem may be found in literature. Just to  mention a few of them we recall the
Hamiltonian approach and the symplectic methods
\cite{Anco,ADM,Waldsymp,Nester,HawHun,Kij,RT},
the  Hamilton--Jacobi--based  techniques
\cite{Booth,Mann94,BY,BLY},
and  the formulations which are directly  based on 
field equations \cite{EM,Silva}.}

In this paper, following a thermodynamical--inspired approach \cite{BY,MannGa,GH1,MannArea}, 
we suggest that the anomalous factor of  the Komar potential is, actually, 
by  no means  ``anomalous''. On the contrary what  is anomalous is the interpretation of
Mass/energy we attribute to it. Indeed the Komar potential turns out to be the right
candidate to describe another kind of gravitational energy,  which we refer to  as  the
\emph{gravitational heat}. 

 The theory of classical thermodynamics can be of great
help  in understanding   this issue. A thermodynamic system is
endowed with  an  internal energy $E$.  Along 
a (reversible) thermodynamic process the internal energy can be
converted into  mechanical work $L$
and into heat $Q$ according to  the first law of thermodynamics $\delta
E=\delta Q-\delta L$.  The amount of energy available to do work in a
reversible transformation (with  initial and final temperature both
equal to $T$) is the free energy
$F$, while the difference
$E-F=TS$ ($S=$ entropy)  is that part of energy which inevitably
transforms into heat; see \cite{Fermi}. 

In the last thirty years a close relationship
between the law of thermodynamics and black hole mechanics  has been
established, initially only on  a formal basis by Bekenstein and Hawking
(see
\cite{Carter,Beckenstein,Hawclass}) and subsequently on a more physical ground by Hawking's
discovery of the black hole radiation; \cite{GH1,Hawquant}. 
Therefore, nowadays we have convincing arguments to state that  black
holes are truly thermodynamic systems (see \cite{Waldreview} for a review on the matter).
As such, we should  be able to attribute to them different kinds of energies, such as the
internal energy, free energy and gravitational heat. According to this
perspective,  the gravitational  entropy was calculated   in \cite{MannArea} from
the gravitational heat, suitably  defined as the difference between the
internal energy (obtained via a Brown--York procedure from an  action
functional  with boundary counterterms) and the free energy defined
semi--classically  as the value on shell   of the action functional itself.

In a completely classical (and therefore different) framework   
 we shall give here a  geometric definition of
internal energy and gravitational heat starting from a unique  generating
formula which relates the variation of the Hamiltonian with the
(pre--)symplectic form  of standard General Relativity. Internal
 energy variations
$\delta E$ will be obtained from the variation on--shell  of the
Hamiltonian. When we  rewrite $\delta E$ in a covariant form it splits
into two contributions. One of them, namely the variation of the Noether
potential,  is the only one surviving once 
Neumann boundary conditions on $\delta E$ are imposed; it then corresponds to
the Komar  energy. The second contribution entering into  
$\delta E$ is instead the so--called covariant ADM correction term
\cite{Remarks,BYOur,CADM,Wald,Wald95}.

When dealing
with stationary spacetimes admitting (observer--dependent) horizons
(hereafter we shall be more precise on the notion  of horizon)  the Komar energy 
turns out to be the gravitational heat
$T S$ (for suitable  geometrically defined  quantities $T$ and $S$) while
the covariant ADM correction term is related to the free energy of the
system. 

The idea to relate somehow  entropy with Noether charges is
surely not new; see e.g.\
\cite{Wald} (see also \cite{Brown} where the same idea was worked out within path integral
methods). We stress however that our definition of entropy  differs from the original  one
due to Wald and it is closest in spirit with Mann's prescription
\cite{MannCre,MannGa}. In this respect  we stress that, even if throughout the paper we
shall deal  with the Noether charge, nowhere in the text we shall make in fact use of
Noether theorem.

We also stress that  the identification of the
Komar energy with the gravitational heat provides    just a geometric
definition of entropy with no rigorous statistical and/or thermodynamical
 interpretation (even though it turns out to coincide with its
semi--classically formulation based on  path--integral techniques; see
\cite{Brown,BY,Wald95}). Nevertheless it is supported  by a great  number of
applications. Just to mention the most trivial of them let us consider a
Schwarzschild black hole of mass M. Its temperature $T$  is equal to
$1\over 8\pi M$ while its entropy $S$ (in fundamental units $
G=\hbar=c=k_B=1$) is
$4\pi M^2$, namely one--quarter of the horizon area. Hence the
gravitational heat
$TS$ turns out to be $M\over 2$ which is exactly the Komar superpotential
computed on any surface enclosing the horizon (and this explains the
``anomalous'' factor ${1/2}$). This -- only  apparently -- lucky coincidence
extends  to less trivial   black hole solutions  and, more
generally,  to  all stationary spacetimes  admitting \emph{any number}  of
 causal horizons \cite{GH2,Parentani,Pad031} (including, beyond stationary black hole
horizons, also cosmological  and Rindler accelerated horizons, i.e.
horizons  which are not associated  with black holes). Remarkably enough
the suggested definition of gravitational heat allows one  to simply compute
the entropy commonly attributed to Taub--Bolt and Taub-nut solutions
\cite{MannLo,TaubBolt,MannGa,Hunter,MannMS,MannArea}, whatever   the physical
meaning is of entropy for these solutions. 

According to our point of view entropy turns out to be a
geometric quantity the  cohomological properties of which can be
physically translated into the first law of thermodynamics. In
agreement with \cite{TaubBolt,MannGa,Hunter,MannMS,MannArea}, entropy then arises 
 when  an obstruction exists to globally foliating
spacetime into surfaces of constant time. In other words, entropy turns
out to be closely related to the coordinate singularities of the solution
describing spacetime. Being coordinate singularities a signal that there
exist regions of inaccessibility for
 observers which are at rest in the given coordinate frame, entropy eventually turns
 out to be related to the existence of  regions hidden to the
observers. This perspective  is close in spirit with 
Padmanabham's recent and interesting work  relating entropy with  
 unobserved degrees of freedom; see \cite{Pad031}. In agreement with 
Shannon's original formulation of the theory of information,
entropy  encodes thence   the information content hidden  beyond
inaccessible regions. Not surprisingly our definition of entropy  
agrees with the definition  given in \cite{Pad031} for static spacetimes,
thus endowing  Padmanabhan's physically--motivated construction with a
mathematical  ground and a much broader domain of applicability.\\

The paper is organized as  follows.
 Starting
from Hamilton's field equations, in section \ref{section2}   we derive the relevant
expression for the variation $\delta E$  of the energy. We  rewrite $\delta E$  into  an
explicit covariant form and, afterwards, we  focus our attention on boundary conditions.
Depending on the choice made between   Dirichlet or Neumann  boundary conditions we obtain
different definitions of quasilocal energies, i.e. \emph{Dirichlet energy} or \emph{Neumann
energy}.  The physical interpretation of
such boundary conditions  is carried out in section \ref{ECH}. It is here recognized that,
in spacetimes admitting causal horizons,  the Komar energy 
is related to  the gravitational heat, while the ADM correction term plays the role of the
free energy. Moreover, the cohomological properties the variational formula for  $\delta
E$ is endowed with allow one to relate the entropy to homological obstructions in  spacetime.
We generalize this idea in section \ref{section4}. We suggest, in complete agreement with
\cite{MannArea}, that entropy arises whenever an obstruction exists in foliating spacetime
into space $+$ time, thus corroborating previous results with an alternative  mathematical
viewpoint. Moreover, the interplay between boundary conditions and the choice of a reference
background is discussed. As a good example,  in section \ref{AdSTB} we  analyse the AdS
Taub--bolt solution and we reproduce within our formalism the same numerical values found
elsewhere by means of
 different techniques. 

The appendix \ref{Appendix A}  deals with spherically symmetric
 solutions with a cosmological constant (Schwarzschild--deSitter)  or with generic
external matter fields. This appendix is introduced  in order to
test  on a simple model   the
main formulae   introduced throughout the paper, thus endowing them  with a direct physical
interpretation. Finally, in  appendix \ref{Appendix B} we carry  out the analogy between
the classical geometric  framework developed in this paper and the semi--classical
statistical approach based on path integral techniques.

\section{The Generating Formula and Boundary Conditions}\label{section2}
Let us consider, in a Lorentzian spacetime $M$ of dimension $4$,  a region 
$D$  that is diffeomorphic  to the
product $\Sigma\times \Re$   where
$\Sigma$ is a 
$3$--dimensional   manifold  with a sufficiently regular boundary $
B=\partial \Sigma$. \footnote{Even though the formalism can be generalized to 
any spacetime dimension
$>2$, for the sake of
simplicity we assume $\dim
M=4$. }
We denote this diffeomorphism by
\be
\psi:\Sigma\times \Re\ram D
\ee
 For any 
$t\in \Re$  a hypersurface  $\Sigma_t\subset D$ is induced by $\psi$
according to the rule  $\Sigma_t=\psi (\Sigma\times
\{t\})=\psi_t(\Sigma)$ and we require  it to be a (p"ortion) of a spacelike
Cauchy hypersurface.  The set of all $\Sigma_t$, for $t$ in
$\Re$,  defines a \emph{foliation}  of $D$ labelled  by the time
parameter  $t$. Moreover, each $\Sigma_t$ intersects the
boundary $\partial D$ in a   $2$--dimensional surface 
$B_t$ which is diffeomorphic  to $B$, for all $t$ in $\Re$. The
diffeomorphism is established  by the map
$\psi_t: B\ram B_t$. Hence the boundary  $\B=\partial D$  is a
timelike hypersurface  globally diffeomorphic to the product 
manifold $B\times \Re$ 
and, physically speaking, it describes
the histories of the observers located on $B_t$. The \emph{time evolution  field}
$\xi$ in
$D$ is defined through the (local) rule $\xi^\mu \nabla_\mu t=1$ and, on
the boundary  $\B$, it is tangent to the boundary itself.
We denote by $u^\mu$ the future directed  unit normal  to
$\Sigma_t$ and we denote by $n_\mu$  the outward pointing
unit normal of $B_t$ in $\Sigma_t$.
For simplicity we assume the foliation to be \emph{orthogonal} so that the vector
$n^\mu$ is also the unit normal of $\B$ in spacetime $M$. Accordingly,
everywhere on $\B$ it holds true that
$\left.u^\mu\,n_\mu\right\vert_{\B}=0$.

 The evolution vector field can be decomposed as
\be
\xi^\mu=N\,u^\mu+N^\mu\label{deti}
\ee
 where $N$ is the lapse and the shift vector $N^\mu$ is tangent  to
the hypersurfaces $\Sigma_t$, i.e.
$ N^\mu\,u_\mu= 0$ and
$\left. N^\mu\,n_\mu\right\vert_{\B}= 0$. The metrics
induced  on $\Sigma_t$, $\B$ and $B_t$  by the metric  
$g_{\mu\nu}$  are given, respectively, by:
\ba
&&h_{\mu\nu}=g_{\mu\nu}+u_\mu \,u_\nu\label{33}\\
&& \g_{\mu\nu}=g_{\mu\nu}- n_\mu \, n_\nu\label{34}\\
&&\s_{\mu\nu}=h_{\mu\nu}-n_\mu \,n_\nu= \g_{\mu\nu}+
u_\mu
\, u_\nu\label{35}
\ea
The metrics \form{33}, \form{34} and \form{35}  with an index raised 
through the contravariant metric $g^{\mu\nu}$ define the projection
operators  in the corresponding surfaces.
 We also denote by 
\ba
&&K_{\mu\nu}=-h^\al_\mu \nabla_\al u_\nu\\
&& \Theta_{\mu\nu}=- \g^\al_\mu \nabla_\al 
n_\nu\\
&&{\cal K}_{\mu\nu}=-\s^\al_\mu D_\al n_\nu
\ea
the extrinsic curvatures, respectively, of  $\Sigma_t$ in $M$, 
of $\B$ in $M$ and  of $B_t$ in $\Sigma_t$.
The symbol $D$  denotes   the
(metric) covariant derivative on $\Sigma_t$ compatible with
$h_{\mu\nu}$.
We shall denote  by $K=K_{\mu\nu} h^{\mu\nu}$,
$ \Theta= \Theta_{\mu\nu}  \g^{\mu\nu}$ and 
${\cal K}={\cal K}_{\mu\nu} \s^{\mu\nu}$ the traces of the
appropriate extrinsic curvatures. The \emph{momentum}
$P^{\mu\nu}$ of the hypersurface $\Sigma_t$ is defined as:
\be
P^{\mu\nu}={\sqrt{h}\over 2\kappa}\left( K
h^{\mu\nu}-K^{\mu\nu}\right)
\ee
while for the \emph{momentum} $\Pi^{\mu\nu}$ of the hypersurface
$\B$ we set:
\be
\Pi^{\mu\nu}=-{\sqrt{\g}\over 2\kappa}\left( \Theta
\g^{\mu\nu}-\Theta^{\mu\nu}\right)\label{Pi}
\ee

Let us consider the  projection of Einstein's
equations (in vacuum or with a cosmological constant) into their normal and tangential
components relative to the leaves $\Sigma_t$ of the foliation: 
\ba
&&\H=0\label{HE1}\\
&&\H_\al=0\label{HE2}\\
 &&\pounds_\xi
h_{\mu\nu}=[h_{\mu\nu}]\label{HE3}\\
 &&\pounds_\xi
P^{\mu\nu}=[P^{\mu\nu}]\label{HE4}
\ea
where $\H, \H_\al$ are the usual \emph{Hamiltonian constraints} \cite{Gravitation},  
$[h_{\mu\nu}]={2\kappa N\over \sqrt{h}}(2P_{\mu\nu}-h_{\mu\nu} P)+2D_{(\mu}N_{\nu)}$
 and $\left[P^{\mu\nu}\right]$ shortly denote the right hand
side of
 the
Hamilton equations, the detailed expression of which is of no
primary interest here and can be found  in
\cite{Gravitation}.
Notice that 
\form{HE3} is nothing but the definition of the momentum $P^{\mu\nu}$
coniugated to the spatial metric $h_{\mu\nu}$, i.e. the Legendre
transformation, while 
equations \form{HE1}, \form{HE2} and \form{HE4} 
correspond, respectively,  to Einstein's equations  
$G^{\mu\nu}u_\mu u_\nu$, 
$G^{\mu\nu}h_{\al
\mu}u_\nu$
and $G^{\alpha\beta}h^\mu_\alpha h_\beta^\nu$  written in a first--order
formalism  in terms of the conjugated fields  $h_{\mu\nu}$ and
$P^{\mu\nu}$.

Let us now denote by $X=\delta g_{\mu\nu} {\partial\over \partial
g_{\mu\nu}}$ a vertical vector field in the configuration bundle of the
Lorentzian metrics (in the sequel we shall shortly denote with $\delta y$
the variation induced on a generic metric--dependent field $y$ by the
variation $X$ of the metric itself). 
If we  multiply the  equations \form{HE1}--\form{HE4} with, respectively,  $\delta N$,
$\delta N^\alpha$,  $\delta P^{\mu\nu}$ and $\delta h_{\mu\nu}$, we sum up all the
obtained expressions and we suitably rearrange the terms,  we finally obtain that
 the  Hamiltonian generating
equations
\form{HE1}--\form{HE4}
is constrained by the \emph{variational formula}:
\be
\delta_X H(\xi, \Sigma_t)=\Omega(\Sigma_t,
X,\pounds_\xi g)\label{analogy}
\ee
where:
\be
\delta_X H(\xi, \Sigma_t)=\int_{\Sigma_t}\left\{\H\, \delta N
 + \H_\mu\,\delta N^\mu +[h_{\mu\nu}] \delta P^{\mu\nu}-[P^{\mu\nu}]
\delta h_{\mu\nu}\right\}d^{3}x\label{deltataH}
\ee
denotes the variation of the Hamiltonian along $X$, while 
\be
\Omega(\Sigma_t,
X,\pounds_\xi g)=\int_{\Sigma_t}
\omega (X, \pounds_\xi g)=\int_{\Sigma_t}
\left\{(\pounds_\xi h_{\mu\nu})\,\delta P^{\mu\nu}
-(\pounds_\xi P^{\mu\nu})\,\delta
h_{\mu\nu}\right\}\,d^{3}\,x
\ee
is the \emph{(pre--)symplectic form}; see  \cite{Waldsymp}.
After a lenghty calculation (see, e.g.
\cite{Booth,BLY}), expression \form{deltataH} can be
conveniently rewritten as:
\ba
\delta_X H(\xi, \Sigma_t)&=&\delta_X\int_{\Sigma_t}\left\{N \H
+N^\al\H_\al\right\} d^{3} x\label{444}\\
&&+\int_{\partial \Sigma_t}d^{2}x \left\{
 N\delta (\sqrt{\s}\, \epsilon)-
N^\al\delta(\sqrt{\s}\, j_\al)
+{
N\sqrt{\s}\over2} s^{\al\b}
\delta \s_{\al\b}\right\}\nonumber
\ea

where 
\ba
&& \epsilon={1\over \kappa} {\cal K},\qquad \qquad \qquad
\qquad \qquad\qquad\qquad \left( \kappa={1\over 8\pi}\right)\\
 && j_\al=-{2\over
\sqrt{
\g}}\,
\s_{\al\mu} 
\Pi^{\mu\nu}\,
 u_\nu\\  && s^{\al\b}={1\over \kappa}\left[( n^\mu 
a_\mu)
\s^{\al\b}- {\cal K}\,\s^{\al\b}+ {\cal K}^{\al\b}\right]\quad
( a_\mu= u^\nu\nabla_\nu  u_\mu)
\ea
are, respectively,  the \emph{surface  quasilocal energy}, the
\emph{surface  momentum}  and the \emph{surface stress
tensor} which  describe the stress energy--momentum content of
the gravitational field inside $B_t$ (see \cite{Booth,Mann94,BY,BLY}
 and references quoted therein). We point out that for non--orthogonal
foliations of spacetime, i. e. $\left. u^\mu\,
n_\mu\right\vert_{\B}\neq0$, an extra term has to be added in \form{444}
which  corresponds to a symplectic boundary structure; see
\cite{Booth,BLY,forth,HawHun,Kij}.

We define the (variation of the) energy 
contained in a region $\Sigma_t$ enclosed by a surface $B_t$
 as the value on--shell  of
the (variation of the) Hamiltonian. From \form{444} we 
obtain:
\be
\delta_X E(\xi,
B_t)=\int_{B_t}d^{2}x \left\{
 N\delta (\sqrt{\s}\, \epsilon)-
N^\al\delta(\sqrt{\s}\, j_\al)
+{
N\sqrt{\s}\over2} s^{\al\b}
\delta \s_{\al\b}\right\}\label{deltaE}
\ee
where now the vector field  $X=\delta g_{\mu\nu} {\partial\over \partial
g_{\mu\nu}}$ is meant to be a solution of the linearized field equations
or, equivalently, it can be viewed  as a  vector tangent to the space of
solutions (which in  turn can be identified with the phase space once
the suitable gauge reductions have been performed; see \cite{Waldsymp}).
Moreover from \form{analogy} we have the on--shell relation:
\be
\delta_X E(\xi,
\partial \Sigma_t)=\int_{\Sigma_t}
\omega (X, \pounds_\xi g)\label{deltaomega}
\ee
Hence,  for stationary spacetimes which satisfy
vacuum Einstein's field equations (possibly with a cosmological
constant), the $2$--form that  defines the variation of
energy is a closed form. Indeed,  the right hand side of
\form{deltaomega} is vanishing when $\xi$ is a Killing
vector: 
\be
\delta_X E(\xi,
\partial \Sigma_t)=0 \quad \hbox{if }\quad  \pounds_\xi g=0\label{2345}
\ee

We stress that, once we fix $B_t$ in the expression
\form{deltaE}, there exist many ``energies'', depending both on the
symmetry vector field
$\xi$ and on the variational vector field $X$; see,
e.\ g.\ \cite{HIO,Anco,Barnich,Nester,forth,Kij}. The former vector determines, via its flow
parameter, how the   observers located on $B_t$ evolve as the ``time'' flows. The
vector field
$X$   fixes instead the boundary conditions on the dynamical fields (e.g.
Dirichlet or Neumann boundary conditions). Each choice of the pair
$(\xi,X)$ gives rise to a particular realization of  a physical system
endowed thence with its own  ``energy content''. 

\begin{Remark}\label{Remark}{\rm
We stress that, despite the fact that the Hamiltonian formulation breaks down  the
covariance of the theory  through the spacetime foliation into
space$+$time, covariance can be nevertheless  restored at the end. Indeed 
formula \form{deltaE} can be conveniently rewritten in an
explicit covariant form as follows:
\be
\delta_X E(\xi,
B_t)=\int_{B_t} {1\over2} U^{\alpha\beta}(\xi,X) ds_{\alpha\beta}\qquad
(ds_{\alpha\beta}=i_\beta\rfloor i_\alpha\rfloor d^4 x)\label{233}
\ee
where 
\be
U^{\alpha\beta}(\xi,X)=\delta \left [{\sqrt{g}\over \kappa}\nabla^{[\beta}
\xi^{\alpha]}\right] + {\sqrt{g}\over \kappa}\, g^{\mu\nu}\,\delta
u^{[\beta}_{\mu\nu}\,\xi^{\alpha]}\label{potcov}
\ee
with $u^{\beta}_{\mu\nu}=\Gamma^{\beta}_{\mu\nu}-\delta^\beta_{(\mu}
\Gamma^\rho_{\nu)\rho}$. The first term in the right hand side of
\form{potcov} is nothing but the variation of the Noether
superpotential $U_{\Kom}(\xi)$, namely the Komar potential,
while the second term is the so--called \emph{covariant ADM correction
term} $U_{\CADM}(\xi,X)$ (see \cite{Remarks,CADM,EM,forth}). 

 Written in terms of variables
adapted to the foliation they read, respectively, as follows \cite{BYOur,forth}\footnote{We
point out that, if we had not assumed the foliation to be orthogonal,   there would appear
additional terms in the right hand side of formulae
\form{U1} and \form{U2} which however vanish when  we assume $\xi$ to be a Killing vector.
With the  assumption of orthogonal boundaries these extra terms are nevertheless vanishing
without any Killing requirement; see
\cite{forth,Wald95}.}:
\be
\delta_X \int_{B_t} U_{\Kom}(\xi)=
{1\over
2\kappa}\int_{B_t}d^{2} x\,\delta\left[ 2\sqrt \s  u^\mu
 \Theta ^\al_\mu
\xi_\al\right]\label{U1}
\ee
and
\ba
\int_{B_t} U_{\CADM}(\xi,X)
&=&-\int_{B_t}\!\! d^{2} x \;\g_{\mu\nu}\, \delta
\Pi^{\mu\nu}\nonumber\\
&=&{1\over
2\kappa}\int_{B_t}\sqrt{\s} d^{2} x  N\left[ 2\delta 
\Theta + \Theta^{\mu\nu} \delta 
\g_{\mu\nu}\right]\label{U2}
\ea
Taking into account the useful formulae \cite{BY}:
\ba
&&\Theta_{\mu\nu}={\cal K}_{\mu\nu}+ ( n^\al 
 a_\al)  u_\mu  u_\nu+2\s^\al_{(\mu} u_{\nu)}
 K_{\al\b}  n^\b\label{split}\\
&&\delta \g_{\mu\nu}=-(2/ N)  u_\mu u_\nu\delta
 N -(2/ N)\s_{\al(\mu}  u_{\nu)}\delta  N^\al +
\s^\al_{(\mu}\s^\b_{\nu)}\delta \s_{\al\b}
\ea 
it easily follows that the sum of \form{U1} and \form{U2} corresponds to 
\form{deltaE}.

We just stress that it is indeed the splitting 
\be
\delta_X E(\xi,
B_t)=\delta_X \left\{\int_{B_t} U_{\Kom}(\xi)\right\}+\int_{B_t}
U_{\CADM}(\xi,X)\label{splitcov}
\ee
 which will
come naturally into play when attempting to describe the gravitational
system in term of thermodynamical variables.
Indeed the two contributions  in \form{splitcov} will be identified,
respectively, with the $\delta (TS)$  and $\delta F$ contributions  to the
internal energy $E$ (where $F$ denotes the  \emph{Helmholtz potential}).

 We also stress that
\form{splitcov}  was obtained in \cite{Waldsymp,CADM,EM,Wald} in a purely
Lagrangian framework as the variation of the (covariantly  conserved) charge associated to
the
\emph{generator of symmetries} $\xi$. From the explicit covariance of
\form{splitcov} it turns then   out that the variation of energy depends
on the generator of symmetries
$\xi$, namely it depends on the observer, but  it does not depend on  the
coordinate system.
\CVD}
\end{Remark}

We point out that the expression  \form{deltaE} can sometimes be integrated in
specific applications,  providing us with  an explicit value of $E$; see e.g. section
\ref{AdSTB} and Appendix \ref{Appendix A}. 
Nevertheless there is no guarantee in general that
there really exists a function $E$ such that its variation fulfills
\form{deltaE}, even if the infinitesimal quantity $\delta_X E$ is always well defined  in our
hypotheses.

 The
problem we  address now is  to implement  boundary conditions which allow
formal integration of expression
\form{deltaE}.\\  

{\noindent \bf Dirichlet  energy}. We define the \emph{Dirichlet
quasilocal  energy}
$\E(\xi,B_t)$ contained in the surface $B_t$  to be the value obtained from
\form{deltaE} by imposing the Dirichlet boundary conditions $
\left.\delta  \g_{\mu\nu}\right\vert_{\cal B}=0
$. 
Since $\left.\delta  \g_{\mu\nu}\right\vert_{\cal B}=0$ implies
$\left.\delta  N\right\vert_{\cal B}=0$, $\left.\delta
 N^\al\right\vert_{\cal B}=0$ and $\left.\delta 
\s_{\mu\nu}\right\vert_{\cal B}=0$, formula \form{deltaE} can be
explicitly integrated:
\be
\E(\xi,B_t)=\int_{B_t}d^{2}x \sqrt{\s}\left(
 N\,\epsilon-
N^\al\,  j_\al\right)+\E_0
\label{e-e0}
\ee
We stress that  we are  integrating in the space of solutions
along a curve   of solutions (i.e. the integral curve of $X$ through
the point $g$) satisfying the boundary Dirichlet conditions. Therefore the
constant of integration 
$\E_0$ can be considered  as the quasilocal energy associated to a
point  $g_0$ in the curve and can be fixed as  the
reference point (or zero level) for the energy \cite{forth}, namely
\be
\E(\xi,B_t)=\int_{B_t}d^{2}x \sqrt{\s}\left\{
 N\,(\epsilon -\epsilon_0)-
N^\al\, ( j_\al-j_{0\al})\right\}
\label{e-e0-2}
\ee
where the subscript  $0$ refers to the background solution $g_0$.
Comparing \form{e-e0-2} and \form{deltaE} we obtain the relevant formula
\be
\delta_X E(\xi,B_t)=\delta_X \E(\xi,B_t)+\int_{B_t}\Pi^{\mu\nu}\,
\,\delta \gamma_{\mu\nu}\, d^2x\label{311}
\ee
which obviously reduces to $\delta_{[D]} E=\delta_{[D]} \E$ when we
consider variations generated by a vector field $X$  satisfying the
Dirichlet boundary conditions $\delta_{[D]}\gamma_{\mu\nu}=0$.

We
remark that formula
\form{e-e0}  is nothing but the value on--shell of the Hamiltonian
ensuing from the canonical reduction of the Trace-K Lagrangian
\cite{BLY,BYOur,Haw-Hor,York}. 

When the observers located on $B_t$ evolve with velocity $\xi^\mu=u^\mu$(i.e. $N=1$,
$N^\mu=0$), the Dirichlet energy \form{e-e0-2} reduces to the Brown--York 
quasilocal energy $\E_{BY}(\xi,B_t)=
\int_{B_t}d^2x \sqrt{\s}
(\epsilon- \epsilon_0 )$; see \cite{BY}.
The 
relationships between the energy
\form{e-e0-2} and the Brown--York  energy  have been analysed in \cite{Booth,BY,BLY} (see
also
\cite{BYOur}). We also point out that there is not a commonly accepted and preferred
choice for the (variation of) energy of a gravitational
system among 
  $\delta_X\E(\xi,B_t)$,  $\delta_X \E_{BY}(\xi,B_t)$) or $\delta_X
E(\xi,B_t)$. We nevertheless stress that both  the Dirichlet expression 
$\delta_X\E(\xi,B_t)$ and the Brown--York definition 
$\delta_X
\E_{BY}(\xi,B_t)$ are a particular case of the more general prescription \form{deltaE}. In
the sequel we shall make use of this latter definition. Indeed, we remark that in all the
specific solutions analysed so far (including Kerr--Newman, BTZ, Taub--bolt, isolated
horizons solutions)  expression \form{deltaE} always gives rise  to the expected numerical
value; see
\cite{HIO,Remarks,BTZ,TaubBolt} and the appendix \ref{Appendix A}.
\\

{\noindent \bf Neumann energy}.
In deriving expression \form{e-e0} we 
have selected the 
components of the boundary metric
$ \g_{\mu\nu}$  as the symplectic control parameters of the gravitational
system \cite{Nester,Kij}, while the response variables  are 
the components of the boundary momentum $\Pi^{\mu\nu}$, namely
the  quasilocal
surface  energy $\sqrt{\sigma} \epsilon$,   the 
quasilocal surface   momentum $\sqrt{\sigma} j_\al$ and
the surface pressure  $\sqrt{\sigma} s^{\al\b}$.

It was argued in \cite{BYpre,BY} that boundary  conditions in General Relativity 
exactly correspond to boundary conditions  of a  thermodynamical ensemble 
and that dynamical fields  which are conjugated to each other in a
symplectic sense,  can be also considered as being   thermodynamically conjugate.
Accordingly,  different ensembles  can be realized  by exchanging  the
fields  that are  kept fixed  as boundary data  with their respective 
boundary conjugated fields. Under this point of view the variational
vector field $X$ turns out to be a variation  into the ensemble realized
with specified boundary conditions.

In obtaining formula \form{e-e0} we kept fixed the boundary $3$--metric
and we obtained the Dirichlet  energy. The boundary fields which are
symplectically conjugated to the boundary  metric are the boundary momenta
$\Pi^{\mu\nu}$ defined in \form{Pi}. Hence we formally describe the
gravitational system in another    ensemble by taking
the momenta as boundary fixed data. We shall now check if the new
-- Neumann --  boundary conditions lead to a mathematically well--defined 
notion of energy, i.e. we will check whether formula \form{deltaE} turns out
to be  integrable under the conditions $\left.\delta
\Pi^{\mu\nu}\right\vert_{\cal B}=0$.
 That this is indeed the case easily
follows by observing that the covariant ADM potential \form{U2} is
vanishing if the much weaker  condition 
\be
\left
.\gamma_{\mu\nu}\delta
\Pi^{\mu\nu}\right\vert_{\cal B}=0\label{NEUm}
\ee
is imposed and, consequently, the
contribution to the variation of energy comes entirely from the Komar
potential, i.e. 
\be
\delta_{[N]}E(\xi,
B_t)=\delta_{[N]} \int_{B_t} U_{\Kom}(\xi)\label{deltaNE}
\ee
 where now $\delta_{[N]}$
refers to variations generated by a vector field $X$  satisfying the
(weak) Neumann  boundary condition \form{NEUm}. Therefore we end up with a new
energy, say \emph{the Neumann quasilocal energy} $Q$, defined as
\ba
Q(\xi,B_t)&=&\int_{B_t}U_{\Kom}+Q_0\nonumber\\
&=&{1\over
\kappa}\int_{B_t}
\sqrt{\sigma} d^{2}x \left\{N \, n^\mu \,  a_\mu -N^\alpha K_{\alpha\beta}\,
n^\beta\right\}+Q_0
\label{EKomar}
\ea
where in the second equality we made use of \form{deti}, \form{U1} and
\form{split}. The constant of integration
$Q_0$ again refers to the energy of a (background) solution $g_0$
lying in the integral curve of
$X$ through the point
$g$.

Comparing now \form{EKomar} with \form{deltaE} we have:
\be
\delta_X E(\xi,B_t)=\delta_X
Q(\xi,B_t)-\int_{B_t}\gamma_{\mu\nu}\, \delta \Pi^{\mu\nu}\,
\, d^{2}x\label{3114}
\ee
Finally,  from \form{311} and \form{3114} we infer that 
\be
\delta_X \E(\xi,B_t)=\delta_X
\left\{Q(\xi,B_t)-\int_{B_t}\Pi^{\mu\nu}\,
\,\gamma_{\mu\nu}\, d^{2}x\right\}
\ee
i.e., the Dirichlet  quasilocal energy $\E$ and the Neumann quasilocal
energy $Q$ are related by a  boundary Legendre transformation exchanging 
boundary metric with boundary momenta.

We shall now analyse to what extent  we can 
refer to the energy
\form{EKomar}  as the physical \emph{gravitational
heat} of a  gravitational system.
\section{ Entropy of Causal Horizons}\label{ECH}

In this section  we shall deal with  $4$--dimensional stationary spacetimes admitting
one or more causal horizons; see \cite{GH2,Parentani,Pad031}. A \emph{causal horizon} $\H$ is
defined as the boundary of the past  of a (future inextensible) timelike
curve  which represents the observer's worldline. This observer dependent
definition encompasses as a special case all (observer independent)
standard black hole horizons but it also includes observer dependent 
horizons such as the cosmological horizon  as well as the Rindler
horizon  (i.e.  the horizon tested by an accelerated observer in
Minkowski spacetime). In a recent paper
\cite{Parentani} the black hole laws of thermodynamics were extended 
 to the  more general setting of causal horizons by relying on
the notion of local horizon entropy. We claim that the  same local
notion of    entropy can be formulated in terms of the Komar energy
\form{EKomar}, extending in this way   Wald's original definition
\cite{Wald} to a broader context. 

In the previous section we gave  the   definition \form{deltaE} for
the (variation of the) internal energy contained in a spacelike region $\Sigma_t$
 bounded  by a $2$--dimensional surface $B_t$.
 Let us now  assume
that  the spacetime is stationary and that the  region
$\Sigma_t$ contains the cross section $H$ of a causal horizon $\H$.
Generalizing  Wald's original  definition
 for stationary black holes, the horizon entropy
$S$ can be  heuristically defined
for \emph{any} $H$  as that quantity  satisfying the first law of
thermodynamics:
\be
\delta_X E(\xi,
H)= T \delta_X S\label{TDS1}
\ee
If the temperature $T$ can be provided  \emph{a priori} by physical
arguments, thence the function $S$ can be truly
identified \emph{a posteriori} with the physical entropy. Namely, expression \form{TDS1} can
be considered as an algorithm to define a quantity $S$ which, by its own
definition, is related to the variation of energy in order that the first
law holds true. 

Let us therefore recall what  a physically--motivated definition of
temperature  could  be. 
In \cite{GH2} it was shown that causal horizons  in stationary spacetimes
which satisfy Einstein's equations in vacuum or with a cosmological
constant (and possibly with an electromagnetic field) are necessarily
stationary axisymmetric  Killing horizons. 
Let us then  identify the vector field \form{deti} with the vector field
which coincides  on the horizon with the null geodesic generator (if
multiple horizons $\H_i$ are present, each one is endowed with its own
vector $\xi_i$ and two different vectors  differ on any horizon for an
axial Killing  vector, the orbits of which  are closed curves  of length
$2\pi$; see \cite{GH2,Parentani}).
In analogy with the black hole case, the  temperature $T$ of a causal
horizon
$\H$ is identified with  the surface
gravity, i.e.:
\be
T={\vert \kappa_\H\vert\over 2\pi}=\left\vert{ n^\mu\nabla_\mu (N^2)
\over 4\pi}\right\vert_\H=\left\vert{ N  n^\mu a_\mu 
\over 2\pi}\right\vert_\H
\label{temp}
\ee
which is constant on $\H$ \cite{Parentani} and  agrees with the Hawking temperature
for  black hole horizons
\cite{Hawquant} or cosmological horizons \cite{GH2}.
 For spacetimes with a single horizon,  the (global) temperature can
be equivalently defined as  the inverse of the  period
$\beta$ of the Euclidean time. 
Namely, when we approach closely to the
horizon along a radial direction  (let us  suppose
that the metric becomes spherically symmetric as it approaches the
horizon\footnote{A wider class of metrics will be considered in the next
section.}) the complexified metric can be written in its Rindler form:
\be
g\simeq R^2 \left({2\pi\over \beta}\right )^2 d\tau ^2 +dR^2 + r^2(R)
d\Omega\label{metricc}
\ee
where $\tau=i t$ is the complexified time,  $d\Omega$ is the metric of the
unit sphere and 
$\beta =1/T$ is given by
\form{temp}. Moreover
\be
r(R)={\pi\over \beta} R^2 + r_+\label{Rr}
\ee
where we have  denoted by $r_+$ the radial position of the horizon. In
the $\tau - R$ plane, the metric
\form{metricc} becomes then regular on the horizon $R=0$ if the
complexified time $\tau$ has a period $\beta$; see e.g. \cite{BY, Hunter,MannArea,Pad031}.

\begin{Remark}{\rm
In spacetimes with multiple horizons $\H_i$ there is no
 global  notion of time--periodicity. Basically, the neighborhood of each
horizon can be covered by a system of coordinates in which the metric is
regular on the horizon and which provides us with a well--defined 
horizon temperature $T_i=1/\beta_i$. However, in the spacetime regions
covered by more than one coordinate system, different notions of
temperature exist  which in general are incompatible among them; see \cite{Pad031}.
Despite of this, 
the  definition
\form{temp} of temperature still has a
\emph{local}  meaning on each single horizon. \CVD}
\end{Remark}

\begin{Remark}\label{Rem1}{\rm
Let us now consider a surface $\Sigma$ of constant time with an outer
boundary denoted by $B$. Let us also suppose that $\Sigma$
intersects a certain number of horizons $\H_i$ with intersection surfaces
denoted by $H_i$. The (variation of the) internal energy 
contained in $\Sigma_t$ is given  by $\delta_X E(\xi,
B)$. Since we are dealing with stationary spacetimes  and since the
surface $B$ is homological to the union of all $H_i$ (i.e. $B\cup_i H_i=\partial\Sigma $
is a homologic boundary)  expression \form{2345} holds then true.
Therefore  the variation of the internal energy receives a contribution
from each cross section $H_i$:
\be
\delta_X E(\xi,
B)=\sum_i \delta_X E(\xi,
H_i)\label{entri}
\ee
 Each $H_i$ being  endowed with its own energy content and with its own
temperature $T_i$ we expect it features  also  an entropy  contribution
$S_i$ according  to  the \emph{local} first law
\be
\delta_X E(\xi,
H_i)= T_i \, \delta_X S_i\label{TDS}
\ee
Basically, for a spacetime with multiple horizons, where a global notion of temperature
 does
not exist  (namely, the spacetime does not
describe a thermodynamical system in equilibrium), there cannot hold a global first law. 
Nevertheless \emph{many} local first laws hold according to \form{TDS}.

We stress that  expression \form{TDS} is calculated directly on the
horizon independently of what exists outside it. This is a rather desirable
property. Indeed even if the whole gravitational system is not in thermal
equilibrium it is nevertheless expected that an observer  located
near the horizon should be able to measure physical properties of the 
horizon itself without knowledge of spacetime regions 
far away from him. 

We also  point out that the  expression \form{TDS}, as it stands, is suitable
for applications to a broader class of horizons, e.g.  isolated horizons
\cite{AshIH}, the geometric characterization of which  is intrinsically
independent on 
the geometric properties of the spacetime surrounding them.
 In the isolated horizons framework, the whole spacetime
is not even   required to be
stationary. Rather,   it is  just  assumed that no flow of matter/energy crosses the
horizon.  Under these conditions the horizon is \emph{isolated} and it can be considered as 
a thermodynamical  system in equilibrium. Therefore the  laws of thermodynamics can be
generalized also to this class  of horizons; see \cite{AshIH,BoothIH}. In 
\cite{HIO} it was indeed shown how the first law can be geometrically  reproduced by means of
\form{TDS}.\\

We finally point out that if we had made use of the definition
\form{e-e0-2} of internal energy  the above construction
would have failed  to work since
$\E(g,H_i)=0$ (being both the lapse and the shift vanishing on $H_i$).

 \CVD}\end{Remark}

 In the purely Lorentzian sector, we recover the
Lorentzian counterpart of expression \form{metricc} by considering the
class of metrics
\cite{Pad031} which are: 
\begin{enumerate}
\item[i)] static in the given reference frame, 
\item[ii)] have vanishing
lapse   on some (compact)
$2$--surface $H_i$ defined by $N^2=0$ (surfaces  corresponding to 
removable singularities of spacetime) and 
\item[iii)] non vanishing derivative
$\nabla_\al N^2\neq 0$ on $H_i$.
\end{enumerate}
 Near each horizon
$H_i$ we can approximate the lapse  function with its Taylor series
truncated to  the first order,  obtaining in this way  the  metric:
\be
g=- R^2 \left({2\pi\over \beta}\right )^2 dt ^2 +dR^2 + r^2(R)
d\Omega\label{metricclor}
\ee
where $T=1/\beta$ is given by \form{temp} while $r(R)$ corresponds to
\form{Rr}. Namely: the zeroes of $N^2$ determine the horizon radii $r_+$ while
the radial derivatives of $N^2$ fix the temperatures $\beta$. When we
consider  variations $\delta_X g$ along a generic vector field $X$, both
parameters
$r_+$ and
$\beta$ are varied.
\\

For the class of  metrics \form{metricclor} the  boundary
condition \form{NEUm} calculated on the horizon $\{R=0\}$
becomes: 
\be
\left
.\gamma_{\mu\nu}\delta
\Pi^{\mu\nu}\right\vert_{H}=-{\pi r_+^2\over \beta^2}\delta \beta
=\pi r_+^2 \, \delta \,T\label{NEUm1}
\ee

Therefore, the (weak) Neumann boundary condition 
$\gamma_{\mu\nu}\delta
\Pi^{\mu\nu}\vert_{H}=0$ 
is satisfied if $\delta \beta\vert_{_{R=0}}=0$.
Namely, the Neumann ensemble is locally realized by the set of metrics
admitting a causal horizon inside the surface $\Sigma$ with a fixed
temperature
$T={1\over\beta}$. Hence, for vector fields $X$ which describe variation
along the Neumann ensemble, from \form{TDS} we obtain: 
\be
\delta_{[N]}(TS)=T \delta_{[N]}(S)=
\delta_{[N]}E(\xi,
B)=\delta_{[N]} \left(\int_{H} U_{\Kom}(\xi)\right)\label{TSK}
\ee
where in the first equality we made use of the property
$\delta_{[N]}(T)=0$ while the last equality comes from
 expression \form{deltaNE}. The above relation can be formally integrated
as:
\be
T\,S=\int_{H} U_{\Kom}(\xi) + Q_0\label{Q-TS}
\ee
where  $Q_0$ is a constant of integration  and corresponds to the Komar
potential calculated for a (background) solution $g_0$ lying in the
integral curve of the vector
$X$. It  corresponds to the ground state for the calculation of the
gravitational heat $TS$. From now on we shall set the  constant  $Q_0$
equal to zero so  that the reference solution corresponds to the
one for which the horizon $H$ shrinks to zero, i.e. the solution for which
$r_+=0$ (since our  analysis is only local we do not forbid the
background solution to have other horizons elsewhere).
\vspace{.6cm}

 It is now easy to verify
that  the entropy, as defined by
\form{Q-TS},  corresponds to  one--quarter of the horizon   area $A_H$.
Indeed, when we approach the horizon $\H$ the metric assumes its 
form
\form{metricclor}   and expression
\form{EKomar}  becomes:
\be
\int_{H} U_{\Kom}=
{1\over
\kappa}\int_{H}
\sqrt{\sigma} d^2x N \, n^\mu a_\mu \um^{\form{temp}}{1\over 4}T
\int_{H}
\sqrt{\sigma} d^2x=T{A_H\over 4} \label{5050}
\ee
\\
provided that the orientation of the unit normal $n^\mu$ is  chosen
so that
${N  n^\mu a_\mu 
/ 2\pi}$
turns always to be positive.\footnote{For non compact horizons, e.g. Rindler horizons
\cite{Pad031}, formula \form{5050} gives a finite value only if the domain of integration
extends up to a finite spatial region. Despite of this,  the entropy for unit of area  is
always a finite constant and it equals  $1/4$.}

We stress that the definition of entropy as it arises from  \form{Q-TS} and \form{5050} is
clearly observer--dependent. Indeed, conditions i)--iii) listed above may be rephrased as
 follows. Given a set of  observers  evolving with $4$--velocity $\xi^\mu$, let us  choose
an observer--adapted system of coordinates in which the observers themselves are at rest. In
this reference frame the metric coordinate singularities  correspond to homological
boundaries and, according to  \form{5050}, each boundary gives rise to its own  contribution
to the total entropy. However, from a physical point of view, the metric coordinate
singularities correspond to regions of inaccessibility for the observers (one--way
membranes).  Entropy turns then out to be associated  to the observer unseen  degrees of
freedom, namely,  to the regions  which the observers  have no physical access to. According
to \cite{Pad031} we might be tempted to state that entropy measures the information content
beyond the hidden regions and that such  an information is related to one--quarter of
the area of the surface  enclosing the observer--unaccessible regions.
\\

Let us now end this section  with some further  thermodynamically--inspired  consideration. 
Let us consider  again  expression \form{233} and let us evaluate it  on the cross section
$H_i$ of a causal horizon.  According to 
\form{Q-TS} it can be rewritten as:
\be
\delta_X E(\xi,
H_i)=\delta_X (T_i S_i) +\int_{H_i} {\sqrt{g}\over \kappa}\, g^{\mu\nu}\,\delta
u^{[\beta}_{\mu\nu}\,\xi^{\alpha]}ds_{\alpha\beta}\label{500}
\ee
From a thermodynamic point of view, having  formally identified  $E(\xi,
H_i)$ with the internal energy of the gravitational system and the
Noether charge with the $TS$ contribution, it is tantamount to formally
identify the second term in the right hand side of \form{500} with the
variation of the gravitational free (or Helmholtz) energy $F(\xi,
H_i)$:
\ba
\delta_X F(\xi,
H_i)&:=&\int_{H_i} {\sqrt{g}\over \kappa}\, g^{\mu\nu}\,\delta
u^{[\beta}_{\mu\nu}\,\xi^{\alpha]}ds_{\alpha\beta}=-\int_{H_i}\gamma_{\mu\nu}\delta
\Pi^{\mu\nu}d^2 x\nonumber\\
&=&\int_{H_i}U_{\CADM}(\xi,X)
\label{Helm}
\ea
 Accordingly,  expression   \form{500} turns out to be nothing but the Gibbs--Duhem
formula \cite{Hunter,MannArea}:
\be
\delta_X E(\xi,
H_i)=\delta_X (T_i S_i)+\delta_X F(\xi,
H_i)\label{termESF}
\ee
Trying to further enhance   the formal analogy between 
gravitation and thermodynamics  we observe that, in the definition
\form{TDS} of the first law, we are taking into account  variations for
which no ``mechanical work'' is done.  Namely, all the variation of energy
$\delta E$ among nearby solutions corresponds to ``heat'' exchange
$T\delta S$. For  such   (reversible) tranformations, thermodynamics
predicts that  the free energy $F$ must obey the law ${\delta F\over
\delta T}=- S$. This is exactly in agreement with the identification
\form{Helm} and with equation
\form{NEUm1}.
For instance, if  we consider two nearby Schwarzschild solutions (of mass, respectively, $M$
and $M+\delta M$)  the variation of energy $\delta E$  turns out to be
$\delta E=\delta M$ (see appendix \ref{Appendix A}) and it is equipartitioned into variation
of ``heat'' and variation of ``free energy'' according to $\delta M=\delta (M/2)+\delta
(M/2)$. This
splitting corresponds, from a thermodynamical point of view, to the splitting
of $\delta E=T\delta S$ into
$T\delta S=\delta(TS) -S\delta T$. The free energy contribution $\delta(M/2)$
therefore corresponds to the $-S \delta T$ part of the splitting. It \emph{does
not} correspond to mechanical work (no energy can be extracted from the hole in any classical
physical process; see e.g. \cite{Gravitation}) but it is due to the fact that the
transformation is not isothermal (if we change the mass of a Schwarzschild black hole we
also change its temperature according to
$T=1/8\pi M$). Indeed 
$-S\>\delta T =-4 \pi M^2\>\delta (1/8\pi M)  =\delta (M/2)$. 
\\

Further  hints corroborating the viability of our definitions will be
exhibited in the appendix \ref{Appendix B} where the analogies   of the
formalism developed so far with existing literature on the subjet will be
analysed.

\section{Generalized Entropy}\label{section4}

Following the guidelines of 
\cite{MannArea}  and adapting Mann's ideas to our formalism, we shall now try to
generalize the identification of the Noether charge with gravitational
heat to a wider class of situations. We nevertheless stress that our
approach differs considerably from that of \cite{MannArea}. Even though the
latter approach turns out to be more physically motivated we believe
that  our definitions are better suited to handle  specific
applications, e.g. the AdS--Taub Bolt solution,  where comparable results
can be  obtained with considerable less efforts.

Let us then consider a stationary spacetime $M$. Let $\xi=\partial_t$ 
be the Killing vector field 
and let us then denote by $\cup_i \H_i$ the union of the fixed point sets 
of
$\xi$. Any $\H_i$ is   an obstruction  to foliating spacetime 
with surfaces of constant time. Therefore we can consider a
surface $\Sigma$  of constant time the boundary of which is formed  by an
outer boundary $B$ together with the union of all the cross sections $H_i$ of
$\H_i$. If a temperature $T_i$ can be defined for any $H_i$, 
following step by step the same   arguments outlined in Remark \ref{Rem1},
we expect that each
$H_i$ gives rise to a contribution $T_i\delta S_i$ to the internal energy
and therefore a contribution 
$ S_i$ to the total entropy. Notice however that now  our
definition encompasses a broader class of solutions, namely  the
ones  with \emph{any} kind of fixed set points of the Killing vector. For
Lorentzian spacetimes,  it includes any compact causal  horizon but also 
non compact  horizons, such as Rindler horizons or Lorentzian CCBH spacetimes (constant
curvature black holes
\cite{MannCre}). When dealing with Euclidean gravity, bolts, nuts as well as
Misner strings \cite{Misner} are allowed to exist.\\

To calculate the contributions to the entropy arising from each
singularity $H_i$, it is a common procedure to  start from the (local)
relation:
\be
T_i S_i= E(H_i)-F(H_i)\label{rrr}
\ee
relating temperature and entropy with  internal  and  free
energies of each singularity $H_i$. The idea to make use of the above
relation dates back  to Gibbons and Hawking \cite{GH2}. They identified the
internal energy with the Hamiltonian and the free energy with  the
gravitational action, as  defined in formula \form{A61} of appendix \ref{Appendix B}. 
In general both the
Hamiltonian and the action, when calculated on a given solution, give an infinite result.
Therefore one has to introduce  a background solution, suitable matched 
with the original solution, and  subtract off from \form{rrr} the
background action and Hamiltonian. In this way one obtains a finite result
which provides us  the difference of entropy between the solution and its
background. 

Another strategy which   avoids the problems related to 
any background fixing procedure has been developed in \cite{MannArea}. It consists in 
modifying the gravitational action, i.e. the gravitational free energy, by
suitably adding boundary counter terms coming from the conjectured AdS/CFT
correspondence. Gravitational internal energy is then obtained,  via a
Brown--York procedure, from the modified action functional. 

We shall instead face up to the problem following the technique of
section \ref{ECH}. First of all let us notice that if we consider the
infinitesimal version of  \form{rrr}:
\be
\delta(T_i S_i)= \delta E(H_i)-\delta F(H_i)\label{rrr2}
\ee
 the problems of  fixing a suitable background is postponed, in specific
applications, up  to the end of calculations where the background will
resort  as a constant of integration. This
implies a higher practicality since   it is not necessary to select 
from the very beginning the reference solution and to carry on
calculations with it. Indeed,
according to our definitions \form{3114} and \form{Helm}, the relation
\form{rrr2} becomes: 
\be
\delta(T_i S_i)=\delta\left\{\int_{H_i}U_{\Kom}\right\}
\ee
which, after integration, leads to the result
\be
T_i S_i=\int_{H_i}U_{\Kom}+Q_0\label{QOOOO}
\ee
Now, it is the constant $Q_0$ which encodes the information relative
to the background. Different choices of $Q_0$ allow to reproduce many
results exhibited elsewhere. For instance,
$Q_0$ can be set equal to zero for compact horizons. This choice, see 
\form{5050},   leads indeed  to the one--quarter area law for the entropy.
Other choices are nevertheless possible and
strictly  necessary for non compact horizons,  where a background subtraction term is
necessary to obtain a finite result. To this end,   we recall  that, given a
variation
$\delta_X$ induced by a vector $X$  in the space of solutions, the
background has to belong to the integral curve of
$X$ through the original solution. This  means that the
given  solution has to 
 be, in the suitable sense,  continuously deformable into its background, i.e. the solution
and its background  have to lie, at least, in a path--connected region in the space of
solutions. This consideration suggests  how the background has to be chosen in many  of the
applications. The example below will help to clarify this point.
\section{AdS Taub--Bolt solution}\label{AdSTB}

Let us consider the class of metrics of the kind \cite{MannLo,MannArea}:
\ba
&&g=V(r)\left( d\tau +2 N \cos\theta d\phi 
\right)^2+{dr^2\over V(r)}+(r^2- N^2)(d\theta^2 +\sin^2\theta d\phi^2)\label{TB} \\
&&V(r)={r^2+N^2-2mr+(r^4-6N^2r^2-3N^4)\, l^{-2}\over r^2-N^2}\qquad   m,N \hbox{ constant }
\nonumber
\ea
Independently on the parameter $m$,  they are all \emph{asymptotically locally AdS}
\cite{Hunter}:
\ba
&&g=V_\infty(r)\left( d\tau +2 N \cos\theta d\phi 
\right)^2+{dr^2\over V_\infty(r)}+(r^2- N^2)(d\theta^2 +\sin^2\theta d\phi^2)\nonumber \\
&&V_\infty(r)={r^2\over l^2} +\left(1-{5N^2\over l^2}\right) +O(r^{-1})
\label{TBA}
\ea
The asymptotic boundary is a squashed $S^3$, rather than $S^1\times S^3$
\cite{Hunter,Misner},  and the constant $N$ parametrizes  the squashing.
When we  apply the definition \form{deltaE} to the metric  \form{TB},  variations $\delta$
are meant inside the class of solutions with asymptotic behavior \form{TBA}. After some
calculations we obtain:
\be
\delta_X E(\xi,
S_R)={R^2\over R^2-N^2}\delta m\label{deltaER}
\ee
where $S_R$ is a surface of constant radius $R$. In the limit $R\rightarrow\infty$ we have 
\be
\delta_X E(\xi,
S_\infty)=\delta m\label{deltaaam}
\ee
so that we can refer to the parameter $m$ as the total Energy/mass of the solution. Notice
that, even though the quantity $\delta_X E$ is known to be an homological invariant (see
equation \form{2345}) formula
\form{deltaER} depends on the radius. The reason  is easily understood.

One of the  fixed points set of the $U(1)$ isometry 
$\tau$ is the  surface
$r=r_0$  where $r_0$ is a solution of $V(r_0)=0$.  The equation $V(r_0)=0$ allows to
determine the ``mass'' parameter $m$ as a function of $r_0$. From \form{TB} we indeed obtain:
\be
m(r_0)={r_0^4 +(l^2 -6N^2) r_0^2 +N^2(l^2-3N^2) \over 2l^2 r_0}
\label{rrrt}\ee
The fixed set points $r=r_0$  is
called a {\it bolt} and its area is equal to
\be
A(r_0)=8\pi N{3r_0^4 +r_0^2(l^2-6N^2) +N^2(3N^2-l^2)\over r_0 l^2}\label{areabolt}
\ee
 Moreover, the metric \form{TB} has another  coordinate singularity, i.e. another break 
down of the foliation in constant  time surfaces,  due to the presence of a Misner string.
It   is a two dimensional coordinate singularity running along the
$z$-axis (i.e.
$\theta=0,\pi$) from the bolt  out to infinity (see \cite{Misner}). 
The Misner string was recognized in \cite{Misner} to be a removable singularity provided we
fix the period
$\beta$ of the Euclidean time to be $\beta=8\pi N$. In addition,  the absence of a conical
singularity at the  bolt $r=r_0$ determines the constraint \cite{Hunter}:
\be
\beta=8\pi N=\left\vert 4\pi\over V'(r_0) \right \vert
\ee
The above equation admits only two solutions, i.e. $r_0=r_b^\pm$ and $r_0=r_n$, where:
\ba
r_{b}^\pm&=& {l^2\pm\sqrt{l^4 -48 N^2 l^2 +144 N^4}\over 12N}\\
r_n&=&N
\ea
for the values $N\le {(3\sqrt{2} -\sqrt{6}) l\over 12}$ corresponding to a real square root.
This two solutions are  known,
respectively, as the Taub-bolt and the Taub--NUT metrics. By substituting $r_0=r_n$ into 
\form{areabolt}, we easily verify that the NUT has zero area.

From the above discussion it follows  that a surface enclosing the  bolt (or the NUT), i.e.
$S_{(r_0+\epsilon)}$, where $r_0=r_b$ (or, respectively, $r_0=r_n$) is not  homologous to
spatial infinity. Owing to the presence of the Misner string  we also have to consider  two
cones
$C_1=\{\theta=\epsilon,r_0+\epsilon\le r<\infty\}$ and
$C_2=\{\theta=\pi-\epsilon, r_0+\epsilon\le r<\infty\}$ wrapping around the
$z$-axis from the bolt $r=r_0$ up to spatial infinity. Spatial infinity is then homologous
to the union $\Sigma=S_{(r_0+\epsilon)}\cup C_1\cup C_2$.
From the homological property \form{2345} we known that each term in $\Sigma$ gives a
contribution to the total energy \form{deltaaam} and, accordingly, to the total entropy.

We can calculate each separate contribution to the entropy following the definition
\form{QOOOO}.
We obtain 
\be
TS_{\{S_{r_0}\}}=\lim_{\epsilon\rightarrow 0}\int_{r=r_0+\epsilon}U_{\Kom}=T\, {A(r_0)\over
4}\label{667}
\ee
for the bolt contribution and 
\ba
TS_{\{C_1\cup C_2\}}&=&\lim_{\epsilon\rightarrow
0}\left\{\int_{\theta=\epsilon}U_{\Kom}+\int_{\theta=\pi-\epsilon}U_{\Kom}\right\}\label{668}\\
&=&\lim_{r\rightarrow \infty}\left\{
{N^2\over l^2} r+ {N^2(3 N^2r_0+r_0^3-l^2 r_0 +m(r_0)\,l^2)\over
l^2(N^2-r_0^2)}+O(r^{-1})\right\}\nonumber
\ea
for the Misner string contribution (the details of the calculations follows the ones
described in \cite{TaubBolt,MannGa} for the asymptotically locally flat  case $l^2\rightarrow
\infty$). The total entropy $S_{r_0}$ is given by the sum of
\form{667} and
\form{668}:
\ba
S_{r_0}&=&{A(r_0)\over 4}+\beta\cdot \lim_{r\rightarrow \infty}\left(
{N^2\over l^2} r+ {N^2(3 N^2r_0+r_0^3-l^2 r_0 +m(r_0)\,l^2)\over
l^2(N^2-r_0^2)}+O(r^{-1})\right)\nonumber\\
&&+\beta\cdot Q_0\label{totent}
\ea
where $Q_0$ is the constant of integration; see \form{Q-TS}.
If we do not take into consideration  the constant of integration, the total entropy
\form{totent} diverges due to the infinite contribution coming from the Misner string (while
it reproduces a finite value in the limit
$l^2\rightarrow
\infty$). Notice however that the divergent part turns out to be independent on
$r_0$. Therefore if we consider the entropy of the Taub--bolt solution relative to the
Taub-NUT  metric we obtain a finite result and, in the meanwhile, we get rid of the constant
of integration which is common for both the solutions. Indeed both Taub--bolt and Taub-NUT
can be joined, in the space of solutions, by a curve $\gamma$ the points of which all share
the asymptotic form \form{TBA}. This can be done by just varying $r_0$ from $r_n$ to $r_b$.
The constant
$Q_0$ corresponds to the Komar integral calculated on an (arbitrarily) fixed point $ g_0$  on
$\gamma$ \footnote{In particular, diverging terms are eliminated from 
\form{totent} for \emph{any} choice of the background $g_0$ in $\gamma$ (and none of them
corresponds to $Q_0=0$). 
 Nevertheless  background choices other than the Taub--NUT solution turn out  to be, as far
as we know,  meaningless and rather artificial.}.  The relative
entropy of  Taub--bolt  with respect to  the Taub-NUT 
 is then given by:
\be
\Delta S=S_{r_b}-S_{r_n}={2\pi N\over r_b l^2}\left[3 r_b^4+r_b^2(l^2-12 N^2)+2r_b N(6
N^2-l^2)+N^2(l^2-3N^2)\right]\label{end}
\ee
for both the values $r_b=r_b^\pm$.
We remark that the entropy \form{end} satisfies  the first law 
\be
T\, \delta
(\Delta S)=\delta m(r_b)-\delta m(r_n)
\ee 
and it also  agrees with the results of \cite{MannLo}.

\section{Conclusions}
We finally point out  that the thermodynamic analysis we have performed
entirely develops from the definition \form{deltaE} of variation of
energy. We remark that this definition, which perfectly agrees with many covariant 
definitions
obtained elsewhere (see,  e.g.   \cite{Nester,CADM,EM,Wald,Wald95,Kij}), was 
calculated directly from the Hamiltonian equations of motion and  therefore
it does not depend on the specific gravitational Lagrangian $L$ inside the
variational cohomology class
$[L]$ we choose to represent the system, where elements of $[L]$ differ
each other only for the addition of divergence terms. About this, 
let us again point out  that, even if throughout the paper we have dealt with the
Noether charge, nowhere in the text we made use of Noether theorem, which is instead
sensitive to the representative chosen inside $[L]$. The splitting of
$\delta E$ into
$\delta (TS)$ and $\delta F$ depends therefore just on the field content of the
theory (i.e.,  Euler--Lagrange equations) and not on a given   Lagrangian description inside
$[L]$. Only if we try to formally integrate these functions of state with the
appropriate boundary conditions we restore the Lagrangian framework and
 it can be  shown (see appendix \ref{Appendix B}) that standard  (and widely accepted)
results are  reproduced in this way. Nevertheless we stress again that in specific
applications it is more convenient for calculations to consider just
variations of physical observables without performing any   a priori
integration. In this way complicated   calculations  related to  background 
choices and  background matching conditions   are avoided in a first
approach. The background will instead come into play just at the end, in a
more manageable fashion, as a ``constant of integration''.\\

As future developments, we believe that the ideas suggested in this paper are worth being 
extended   to deSitter and Anti deSitter Einstein--Gauss--Bonnet gravity in order
to study  the possibility of negative entropy; see \cite{Odi}. Moreover, the analogy
between Noether charge and gravitational heat deserves to be analyzed in  processes of
 collapse of a star into a black hole in order to understand if a sort of gravitational heat
can be  defined prior  to collapse. 

We would like to thank  S.\ D. Odintsov and R.\ B.\ Mann
for having addressed our attention on these future fields of investigation.

\section{Acknowledgments}
 We are very grateful to R.\ B.\ Mann for a careful reading of this paper and for helpful 
comments and suggestions  on the subject.

This work is partially supported  by GNFM--INdAM (research project ``\emph{Metodi geometrici
in meccanica classica, teoria dei campi e termodinamica}'') and by MIUR (PRIN 2003 on
``\emph{Conservation laws and thermodynamics in continuum mechanics and field theories}'').

\begin{appendix}
\section{Spherically Symmetric Solutions \label{Appendix A}}

Let us consider a static spherically  symmetric solution of the kind:
\be
g=-f(r) dt^2 +{dr^2\over  f(r)} +r^2 d\Omega\label{gss}
\ee
where $d\Omega= d\theta^2+\sin(\theta)^2 d\phi^2$. The explicit
form of $f(r)$ is dictated by Einstein's  field equations and we shall analyze it later on.

The variation of the energy \form{deltaE} contained inside  a region
$\Sigma_t$ of constant time bounded by a a two--sphere $S_R$, i.e.
$B_t:=\{r=R\}$, turns out to be:
\be
\delta_X E(\xi,
S_R)=\int_{S_R}d^2x \,
 N\,\delta (\sqrt{\s}\, \epsilon)=-{R\over 2}\left.\delta f\right\vert_R
\label{deltaEss}
\ee
where $\xi=\partial_t$. 
The Noether charge 
\form{EKomar}, apart for the constant of integration which is set equal to zero throughout
this section,  becomes:
\be
Q(\xi,S_R)={R^2\over 4}  \left\vert f'(R)\right\vert \label{Qss}
\ee
having denoted with a prime the derivative with respect to the
$r$--coordinate. The absolute value arises because when $
f'>0$, e.g. on  black hole horizons,  the normal to the sphere is chosen
to point outward, along  the direction of increasing $r$. When
$f'<0$, e.g. on cosmological horizons (see below), the unit normal  is instead chosen
to point inward toward the direction of decreasing $r$. In this way the term 
${n^\mu\nabla_\mu f
\over 4\pi}= {N  n^\mu a_\mu 
\over 2\pi}$ in the Noether charge \form{EKomar}
turns always to be positive.

 Dirichlet boundary conditions for \form{deltaEss} have been extensively
analyzed in \cite{Mann94,BY}. We instead point out here  that (weak) Neumann boundary
condition  \form{NEUm} corresponds to:
\be
-\int_{S_R} \gamma_{\mu\nu}\delta
\Pi^{\mu\nu}=-{R\over 4}\left\{ R\, \delta
f'(R)+2\, \delta f(R)
\right\}
=0\label{NEUmss}
\ee
It is therefore easy to check that Neumann variations $\delta_{[N]}
Q(\xi,S_R)$ of the Noether charge \form{Qss} give rise to the variation of energy 
\form{deltaEss}. No anomalous factor arises in this case!

Now, let us suppose that $f(r_i)=0$  for some  $r_i$,
($i=1,2,\dots$) and that $f'(r_i)\neq 0$. We define  the temperature
$T_i$ of each horizon $r=r_i$ according to \form{temp}, i.e.:
\be
T_i=\left\vert{\kappa_i\over 2\pi}\right\vert = \left\vert{f'(r_i)\over 4\pi}\right\vert
\ee
On the horizon, \form{NEUmss} is equivalent to $\delta T_i=0$ and the
Neumann energy \form{Qss} becomes:
\be
Q(\xi,H_i)= \left\vert{f'(r_i) \over 4\pi}\right\vert \pi
r_i^2=T_i {A_i\over 4}:=T_i\,S_i\label{QAss}
\ee
where $A_i$ is the area of the horizon's cross section. Therefore the above expression
has the  $T_iS_i$ form if the entropy is identified with $S_i:={A_i\over 4}$. We stress that
so far we have not required the solution to be a vacuum solution (possibly with a
cosmological constant). Therefore the result \form{QAss} holds true for any spherically
symmetric static solution. 

Further considerations require to  consider on separate grounds the cases  in which: 

\noindent i)  the metric \form{gss} is
a solution of Einstein's equations with a (positive) cosmological
constant; 

\noindent ii) the metric  is
a solution of Einstein's equations with external matter fields.\\

{\noindent \bf i) Schwarzschild--deSitter solutions}

If we set
\be
f(r)=1-{2m\over r}-{r^2\over l^2}\label{fdess}
\ee
in the metric \form{gss} we obtain the Schwarzschild--deSitter metric. 
The cosmological constant is $\Lambda=3/l^2$ and $m$ is the mass parameter. If $m=0$ we
recover deSitter spacetime while Schwarzschild  solution is obtained for $\Lambda=0$.

 Provided that
$m<l/\sqrt{27}$ the solution has two positive roots, $r_+$ and $r_{++}$
($r_{++}>r_+$) corresponding, respectively, to the  black hole horizon and the cosmological
horizon. From \form{fdess} we have:
\be
m={r_0\over 2}\left( 1-{r_0^2\over l^2}\right)\label{rmr}
\ee
for both the values $r_0=r_+$  and $r_0=r_{++}$.

When $r$ ranges from $r_+$ up to $r_{++}$ the vector field $\xi =\partial_t$ is
timelike and it can describe the velocities of  observers located between the two horizons
(of course this is just one possible choice; see
\cite{Corichi} for another normalization of $\xi$).

Expression \form{deltaEss} becomes:
\be
\delta_X E(\xi,
S_R)=\delta m
\label{deltaEssDES}
\ee
and does not depend on the radius of integration, as expected from  \form{2345}. 

The temperatures on  $r_+$ and $r_{++}$ are, respectively:
\ba
T_+={f'(r_+)\over 4\pi}={1\over 4\pi} \left( {1\over r_+} -{3 r_+\over
l^2}\right)\label{881}\\
T_{++}=-{f'(r_{++})\over 4\pi}=-{1\over 4\pi} \left( {1\over r_{++}} -{3 r_{++}\over
l^2}\right)\label{882}
\ea
where the minus sign in \form{882} arises since the surface gravity ${f'(r_{++})\over 2}$ is
negative on the cosmological horizon.
 According  to \form{QAss} the integral of the Komar
charge  allows to obtain the one--quarter law  for both the horizons $r_0=r_+, r_{++}$:
\ba
S_{0}=\pi r_{0}^2 \label{entr00}
\ea
Notice from \form{rmr}, \form{881} and \form{entr00} that, on the black hole horizon, 
 the relation
$\delta m=T_+ \delta S_+$ holds true. Owing to \form{deltaEssDES} this
relation corresponds to the first law 
\be\delta E(\xi, r_+)=\delta m=T_+\delta S_+\label{1bh}
\ee
On  the cosmological horizon, on the contrary, it holds true the relation $\delta m=-T_{++}
\delta S_{++}$ so that we expect  $\delta E(\xi, r_{++})$ to be equal to $-\delta m$ if we
want  to recover the first law for the cosmological horizon, too. This result is only
apparently in contrast with  \form{deltaEssDES}. Indeed, in
obtaining 
\form{deltaEssDES} the  spacelike unit normal $n^\mu$ was assumed to point in the direction
of increasing $r$. Nevertheless on the cosmological horizon we have to change the sign, i.e.
\be
\delta_X E(\xi,
r_{++})=-\delta m
\label{deltaEssDESH}
\ee
if we want the unit normal to point toward the observer. In this way the expected value
is recovered and the first law is satisfied.
 Notice that a negative value for the variation of energy contained inside a
cosmological horizon is physically expected; see
\cite{Corichi,Teitelboim}. Indeed, if a test particle of mass  $\delta m$ is thrown across
the cosmological horizon from the interior region, the area of the horizon increases but we
expect that the energy enclosed by the horizon should decrease according to
\form{deltaEssDESH}.\\

Notice that, even though the absolute value $\vert\delta E\vert=\delta m$ between two nearby
Schwar\-zschild--deSitter  solutions is the same when computed on both the
horizons,  the energies of the horizons $E(\xi,
r_{++})$ and $E(\xi,
r_{+})$ can be nevertheless different from each other (even though, in absolute value, they
change for the same amount during the variation). 
From \form{1bh} we obtain:
\be
E(\xi,
r_{+})=\int_0^{r_+} {\delta m(r)\over \delta r} \delta r=m(r_+)\um^{\form{rmr}}{r_+\over
2}\left( 1-{r_+^2\over l^2}\right)
\ee
where the normalization has been reasonably fixed so that the energy vanishes when the
horizon mass vanishes, i.e. the background spacetime is the deSitter solution. Notice that,
for $l^2\rightarrow\infty$  the usual mass ${r_+\over
2}$ of Schwarzschild black hole is
recovered.

According to \form{deltaEssDESH}, for the cosmological horizon we obtain:
\be
E(\xi,
r_{++})=-m(r_{++})+E_0
\ee
In this case different choices are available for the constant of integration. For instance,
if we set
$E_0=0$
 we fix  again the deSitter  background. Another possible choice is
$E_0={2l\over\sqrt{27}}$. In this case $E(\xi,
r_{++})\rightarrow E(\xi,
r_{+})$ in the Nariai limit $m\rightarrow l/\sqrt{27}$ when $r_+$ and $r_{++}$ approach to
each other. We refer the reader to \cite{Corichi} for some deeper insight into the subject
(see also \cite{MannGhe}).\\

{\noindent \bf ii) External matter fields}

If we set
\be
f(r)=1-{r_+\over r}-{1\over r} \int_{r_+}^r \epsilon(\bar r ) \bar r ^2 d\bar r \label{855}
\ee
in the metric \form{gss} we obtain a solution of Einstein's field equations  with external 
matter characterized by  a density
\be
\rho(r)={\epsilon(r)\over 8\pi},
\ee
radial pressure
\be
p(r)=\rho(r)\label{pr}
\ee
together with angular pressures $T^\theta_\theta=T^\phi_\phi={G^\theta_\theta\over\kappa}$,
the explicit form of which is of no interest here; see \cite{HEllis,Padss}. The function
\form{855} has been normalized
\cite{Padss} in order to have an horizon at
$r_+$ with temperature:
\be
T_+={f'(r_+) \over 4\pi}={1 \over 4\pi}\left({1\over r_+} -\epsilon(r_+)
r_+\right)\label{TTT}
\ee
We however do not exclude that  $f(r)$ can be zero for other values  of $r$.

Expression \form{deltaEss} now becomes:
\be
\delta_X E(\xi, S_R)= \delta\left\{ {r_+\over 2 }+  {1\over 2 }\int_{r_+}^R\epsilon(\bar r )
\bar r ^2 d\bar r  \right\}
\label{dEssDES}
\ee
for a generic sphere of radius $R$ and 
\be
\delta_X E(\xi,r_+)= \delta\left\{ {r_+\over 2 } \right\}
\label{asdf}
\ee
for the horizon.

 If we consider variations which keep  the radius $R$ fixed in \form{dEssDES}, we easily
obtain: 
\be
\delta_X E(\xi, S_R)=\delta_X E(\xi,r_+)-{1\over 2 }\epsilon({r_+} )
r_{+}^2 \delta r_+\label{991}
\ee
Since from \form{TTT} we have:
\be
\delta_X E(\xi, S_R)=T_+ \delta S, \qquad S=\pi r_+^2\label{khg}
\ee
by comparing  \form{991} and  \form{khg} we obtain the first law:
\be
\delta_X E(\xi,r_+)=T_+ \delta S + p_{_+} \delta V_{_+}\label{93}
\ee
where $p_+$ is the radial pressure \form{pr} on the horizon, while $\delta V_{_+} =4\pi
r_{+}^2
\delta r_+$. Hence $p_{_+} \delta V_{_+}$ is the work term given by the product $p_{_+} 4\pi
r_{+}^2$ of the radial force multiplied by the radial displacement $\delta r_+$.

Therefore expression \form{93} constitutes the generalization of formula \form{TDS} in
presence of external matter. Namely, the amount  of  energy  to virtually  increase the
radius of the horizon of a quantity  $\delta r_+$ is partitioned into a  part  $T_+ \delta
S$ which comes from the increase of entropy  (due to   the expansion of the horizon area) 
and a work term
$p_{_+}
\delta V_{_+}$ which has to be spent   to contrast the external pressure, see \cite{Padss}.


\section{Comparison with the semi--classical approach \label{Appendix B}}
Let us now analyse the interplay between the formalism developed so far
and the existing literature on the matter.
First of all we stress 
 that the  identification of the Komar energy
\form{EKomar} with the gravitational heat $TS$  ensues from the heuristic
definition of entropy given in \form{TDS}, which a priori 
has no direct physical ground. The identification  is  nevertheless in
agreement  with
 semi--classical statistical approaches,  based on path integral
techniques, where entropy is identified  by the value of the
microcanonical action functional (evaluated  on a
--complexified--  stationary black hole solution; see \cite{Brown,BY,Wald95}).  

Indeed,  let us
consider a closed surface  $B$  surrounding  the horizon $H$  such
that $B\cup H$ is the homologic boundary $\partial
\Sigma$ of a region $\Sigma$. We can  rewrite equation \form{termESF}
as\footnote{For notational convenience, from now on, we shall suppress
the index $i$.}:
\ba
\delta_X (T S )&=&
\delta_X E(\xi,
H)-\delta_X F(\xi,
H)\nonumber\\
&\um^{\form{2345}}&
\delta_X E(\xi,B)-\delta_X F(\xi,
H)\nonumber\\
&\um^{\form{splitcov}+\form{Helm}}&\delta_X\left\{\int_B
U_{\Kom}\right\}+\int_{B}U_{\CADM}(\xi,X)-\int_{H}U_{\CADM}(\xi,X)
\nonumber\\
&\um&\delta_X
\left\{\int_B U_{\Kom}\right\}+\int_{\partial
\Sigma}U_{\CADM}(\xi,X)
\nonumber\\
&=&\delta_X\left\{\int_B U_{\Kom}+\int_{\Sigma} i_\xi\rfloor
L\right\}\label{5353}
\ea
where $L={1\over 2\kappa} \sqrt{g}( R-2\Lambda) \, ds$ is the Hilbert
Lagrangian. In the last equality we made use of Stokes' theorem
together with the following  relations which hold true on--shell for
stationary solutions:
\ba
i_\xi\rfloor\delta L&=&i_\xi \rfloor d \left\{{\sqrt{g}\over 2 \kappa}\,
g^{\mu\nu}\,\delta u^{\alpha}_{\mu\nu}ds_\alpha\right\}\qquad
(u^{\alpha}_{\mu\nu}=\Gamma^{\alpha}_{\mu\nu}-\delta^\alpha_{(\mu}
\Gamma^\rho_{\nu)\rho})\nonumber\\
&=&\pounds_\xi
\left\{{\sqrt{g}\over 2 \kappa}\, g^{\mu\nu}\,\delta
u^{\alpha}_{\mu\nu}ds_\alpha\right\}
-d i_\xi \rfloor\left\{{\sqrt{g}\over2  \kappa}\,
g^{\mu\nu}\,\delta u^{\alpha}_{\mu\nu}ds_\alpha\right\}\nonumber\\
&=&-d \, i_\xi \rfloor\left\{{\sqrt{g}\over 2 \kappa}\,
g^{\mu\nu}\,\delta u^{\alpha}_{\mu\nu}ds_\alpha\right\}\nonumber\\
&\um^{\form{potcov}}&d U_{\CADM}(\xi,X)\label{555}
\ea 
From \form{5353} we obtain (apart from a constant of integration):
\ba
S&=&\beta\left\{ \int_B U_{\Kom}+\int_{\Sigma} i_\xi\rfloor
L\right\}\nonumber\\
&=&\int dt\int_B U_{\Kom}+\int dt\int_{\Sigma} i_\xi\rfloor
L\nonumber\\
&=&\int_{\B}dt\wedge U_{\Kom}+\int_{D} 
L=
I_m
\ea
where $I_m$ is the  microcanonical action functional for a region
$D=\beta\times\Sigma$ of spacetime with outer boundary
$\B=\beta\times B$ ; see, e.g.
\cite{BY,Wald95}. Hence definition
\form{Q-TS} together with the relation \form{termESF}  exactly
corresponds to Brown--York's definition of entropy when  stationary black
holes are considered.\\

We also stress
that the identification $S=\beta \int_{H} U_{\Kom}$ exactly agrees with
the  definition of entropy \cite{Pad031}:
\be
S={\beta\over
\kappa}\int_{H}
\sqrt{\sigma} d^2x N \, n^\mu a_\mu
\ee
 which Padhmanabhan  has given for static
spacetime, as it is can be easily inferred from \form{EKomar}. 
Nevertheless the definition
\form{Helm} of free energy
 differs from Padhmanabhan's definition \cite{Pad031} in which free
energy is identified with the Hilbert Lagrangian. Our definition is
instead more  tied with the Trace-K action functional \cite{BY}. By making
use of the
 relations \form{555}, definition \form{Helm}
can be indeed
conveniently rewritten as:
\be
-\delta_X F(\xi,
H)=\int_\Sigma i_\xi\rfloor \delta  L -\int_{B}U_{\CADM}(\xi,X)
\ee
If we consider the Dirichlet boundary condition $\left.\delta_X
\gamma_{\mu\nu}\right\vert_B=0$ on the outer boundary $B$ (which in many
applications can be identified with spatial infinity), namely, if we
consider the class of metrics which approach  a fixed background
$g_0$ on $B$, the above expression can be integrated. From \form{U2} we 
hence obtain:
\be
-F(\xi,
H)=\int_\Sigma  i_\xi L -{1\over
\kappa}\int_{B}\sqrt{\s} d^2 x  N\left(
\Theta- \Theta_0\right)\\
\ee
After  multiplying the above expression by $dt$ and  integrating over the
time interval $\beta$ we finally obtain the trace--K action functional
$I_g$:
\be
I_g:=\int_D  L -{1\over
\kappa}\int_{\B}\sqrt{\s} d^2 x  N\left(
\Theta- \Theta_0\right)=-\beta \cdot  F(\xi,
H)\label{A61}
\ee
(which coincides, in the hypotheses we are dealing with,  with the first--order
\emph{covariant} Lagrangian for General Relativity; see \cite{BYOur}).
  We recall that,  in a
semi--classical path integral approach, $I_g$ 
corresponds, in the Euclidean sector,   to 
 the logarithm of the thermodynamic partition function;
see
\cite{BY,GH1}.


\end{appendix}


\end{document}